\documentclass{elsart}

\usepackage{graphicx}
\usepackage{epsfig}

\usepackage{amssymb}
\usepackage{color}

\begin{document}

\begin{frontmatter}



\title{Ionization Efficiency Study for Low Energy Nuclear Recoils in Germanium}
\author[usd]{D. Barker},
\author[usd]{W.-Z. Wei},
\author[usd]{D.-M. Mei\corauthref{cor}}, and 
\corauth[cor]{Corresponding author.}
\ead{Dongming.Mei@usd.edu}
\author[usd,tgu]{C. Zhang}

\address[usd]{Department of Physics, The University of South Dakota, Vermillion, South Dakota 57069}
\address[tgu]{College of Sciences, China Three Gorges University, Yichang 443002, China}

\begin{abstract}
We used the internal conversion ($E_0$ transition) of germanium-72 to 
indirectly measure the low energy nuclear recoils of germanium. Together with a 
reliable Monte Carlo package, in which we implement the internal conversion 
process, the data was compared to the Lindhard ($k$=0.159) and Barker-Mei 
models. A shape analysis indicates that both models agree well with data 
in the region of interest within 4\%. The most probable value (MPV) of the
nuclear recoils obtained from the shape analysis is 17.5 $\pm$ 0.12 (sys) $\pm$ 
0.035 (stat) keV with an average path-length of 0.014 $\mu$m. 
\end{abstract}
\end{frontmatter}

\maketitle

\section{Introduction}
Understanding detector response to low energy nuclear recoils is imperative to
the interpretation of experimental results from a detector designed to search 
for WIMPs (weakly interacting massive particles), a dark matter candidate. 
Direct detection of low mass WIMPs occurs in the low energy region of detectors 
with a threshold down to sub keV. Since the threshold energy represents the 
visible energy in the detector, understanding the ionization efficiency of the 
detector response to low energy nuclear recoils is crucial to calculating the 
recoil energy. An example of the need for this requirement is the claim of 
experimental evidence for dark matter by CoGeNT~\cite{CoGeNT2011} that has 
been unverified by CDMS II~\cite{cdms2012}. Both experiments use 
germanium as the target material. Thus, similar results are expected 
if the detection thresholds for both experiments were determined using 
a standardized ionization efficiency, which accurately 
accounts for all processes that occur at a low energy range. This 
standardized ionization efficiency must also be validated with 
measurements in combination with reliable Monte Carlo simulations. 

Two different approaches can be used for modeling ionization efficiency in 
germanium detectors~\cite{lind,dbar}. One model, traditionally used for a 
number of different detector materials, is proposed by 
Lindhard {\it et. al.}~\cite{lind}:
\begin{equation}
\varepsilon = \frac{k\cdot g(\epsilon)}{1+k\cdot g(\epsilon)},
\end{equation}
where $k$ = 0.133$Z^{2/3} A^{-1/2}$, 
$g(\epsilon) = 3\epsilon^{0.15} + 0.7\epsilon^{0.6} + \epsilon$, and
$\epsilon$ = 11.5$E_{r} Z^{-7/3}$ for a given atomic number, $Z$, mass number, 
$A$, and recoil energy, $E_r$. However, this model has not been proved accurate 
at low energies as the theoretical derivation has uncertainties in this 
region~\cite{lind}. 

Another model designed for low energy interactions in germanium was proposed by 
Barker and Mei~\cite{dbar}. This 
model takes into account the fraction of nuclear stopping 
power that contributes to the ionization efficiency at low 
energies~\cite{dbar}. The Barker-Mei model can be expressed as:
\begin{equation}
\varepsilon_{c} = \frac{0.14476\cdot E_{r}^{0.697747}}{-1.8728 + exp[E_{r}^{0.211349}]}.
\end{equation}
This model is valid for recoil energies, $E_{r}$, from 1 keV to 100 keV. 
However, the Barker-Mei model has not been experimentally verified. The purpose of this paper 
is to address that issue. 

\begin{figure}[tb!!]
\includegraphics[angle=0,width=12.cm] {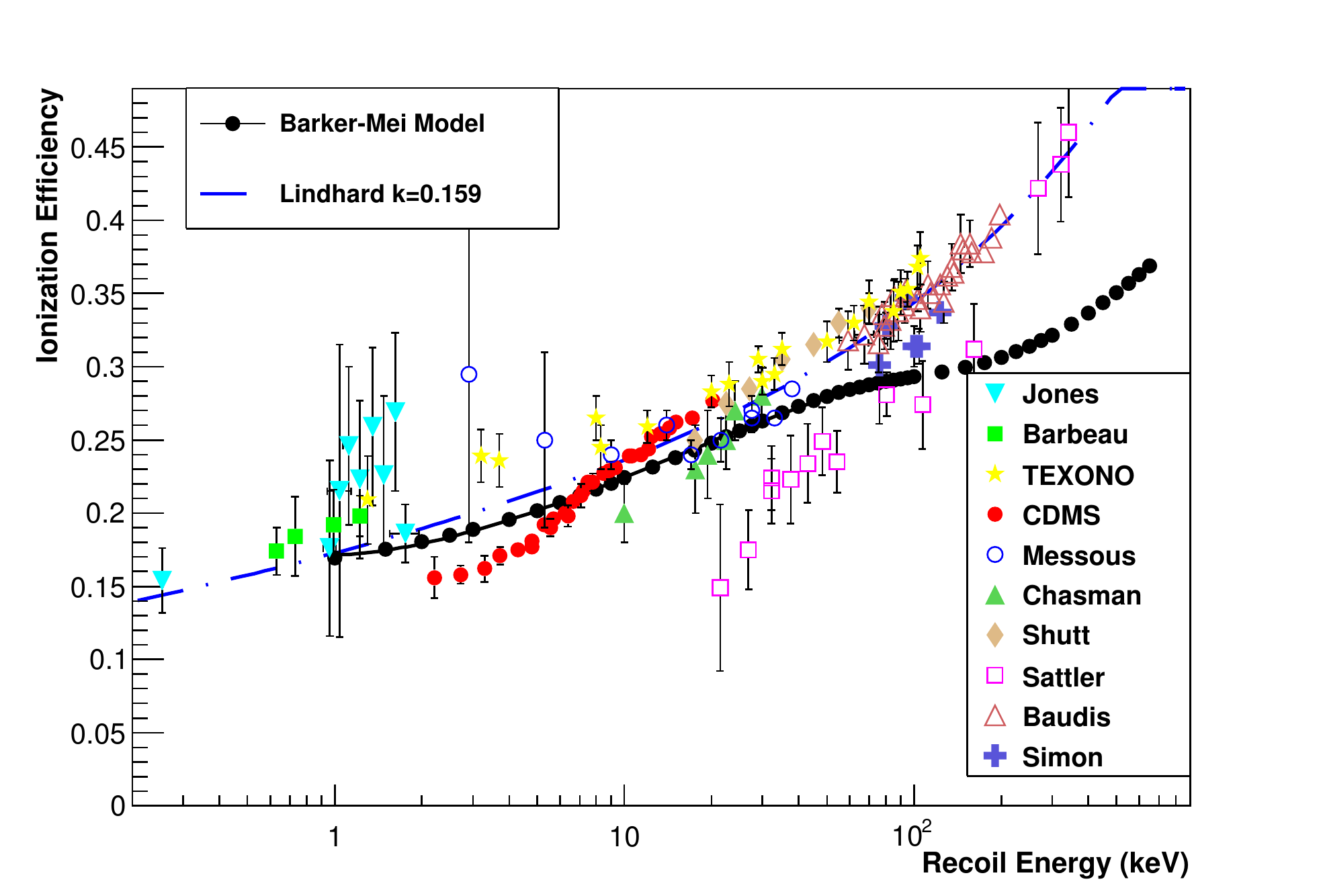}
\caption{\small{The Barker-Mei and Lindhard models of ionization efficiency with the experimental data~\cite{kwj, psb, TEXONO, yme, cch, tsh, ars, lba, esi}.}}
\label{fig:BMLindwdata}
\end{figure}

A comparison between the two models and existing data was 
performed~\cite{dbar}, and Fig.~\ref{fig:BMLindwdata} shows both models 
together with available experimental data.
As shown in Fig.~\ref{fig:BMLindwdata}, the two models agree 
in the low energy region but disagree with the available 
data points in the same region. 
Further validation of the Lindhard and Barker-Mei models, using
more measurements and more accurate Monte Carlo simulation, was necessary. 

Taking exact nuclear recoil measurements can be challenging at lower energies 
when systematic errors are introduced by a variety of sources. For example, 
thermal neutrons and elastic/inelastic scattering have their own 
uncertainties. Utilizing thermal neutrons requires the full 
absorption of out-going gamma rays measured by another detector in coincidence. 
Without this additional measurement, Compton scattering from out-going 
gamma rays within the germanium detector can contaminate the visible energy. 
In a measurement of neutron elastic scattering, a Monte Carlo simulation 
must be incorporated to exclude multiple scatters. It is also necessary to 
precisely measure the scattering angle and time-of-flight of the out-going 
neutrons. With inelastic scattering, the Compton scattering due to 
de-excitation of gamma rays in the detector can contaminate 
the signal. All of these techniques need to be implemented with an accurate 
Monte Carlo that reduces systematic errors in order to obtain a reliable 
ionization efficiency. However, current popular
simulation tools often need to be tuned to simulate inelastic scattering 
processes. 

Therefore, we desired a simple method to accurately measure nuclear recoils in 
germanium. The $E_0$ transition of germanium-72 ($^{72}$Ge(n,n$^{'}$e)), which 
is the internal conversion process for this nucleus, was chosen. The $E_0$ 
transition of $^{72}$Ge(n,n$^{'}$e) is induced when neutrons inelastically 
scatter off a $^{72}$Ge nucleus. After this collision, the $^{72}$Ge nucleus is 
left in an excited state. When the nucleus returns to ground state, it does not 
directly produce a gamma-ray (as is common for other nuclei), but interacts 
electromagnetically with the inner shell electrons and causes one to be emitted 
from the atom~\cite{kenn}. The physics process for the internal conversion of 
$^{72}$Ge can be generalized as:
\begin{equation}
n + ^{72}Ge \longrightarrow n' + ^{72}Ge^{*}
\hookrightarrow ^{72}Ge + e^- + X-ray,
\end{equation}
where $^{72}$Ge$^{*}$ denotes the excited 0$^{+}$ state. The total detectable 
energy from the $^{72}$Ge$^{*}$(0$^{+}$) to $^{72}$Ge(0$^{+}$) transition is 
691.6 keV, which includes energy from the X-ray. This energy is well within the 
visible range of traditional detectors, and both electron and X-ray can be 
detected with a consistency of approximately 100\%. The characteristic 
energy, 691.6 keV, is superimposed with the nuclear recoil of 
a $^{72}$Ge nucleus to form a quasi-triangular shape, which is distinguishable 
from other processes in the spectrum. Thus, we can observe low energy nuclear 
recoils without taking low energy nuclear recoil measurements by extrapolating with
the known quasi-triangular fit. This 
quasi-triangular shape, starting at 691.6 keV, has been studied in great 
detail~\cite{adr,rwor,gfe,gfe2,ege}. The tail of the quasi-triangular shape comes from
a combination of the 691.6 keV energy deposition and nuclear recoil energy due 
to neutron scattering.

Depending on the incident neutron energy and scattering angle~\cite{adr}, 
nuclear recoil energy can be expressed as:
\begin{equation}
E_{r} = \frac{4mME_{n}}{(M+m)^2}(cos^2\theta),
\end{equation}
where E$_{r}$ is the nuclear recoil energy, E$_{n}$ is the neutron energy, M is 
the nucleus mass, m is the neutron mass, and $\theta$ is the scattering angle 
between the incident neutron and the recoil nucleus. The quasi-triangular shape 
is created by adding the energy from the nuclear recoils to the 691.6 keV energy 
deposition: 
\begin{equation}
E = 691.6 keV + \epsilon \cdot E_r, 
\end{equation}
where $E$ is the observed energy and $\epsilon$ is the ionization efficiency. 
The nuclear recoil energies can be determined using a Monte Carlo simulation 
with applied ionization efficiency, if an agreement with the measurements is 
obtained.  

To validate the ionization efficiency models proposed by Lindhard and 
Barker-Mei, a Geant4.9.2-based Monte Carlo simulation package~\cite{geant4},  
corrected for the internal conversion processes, was used. This 
simulation package was verified using a well-calibrated $^{60}$Co radioactive 
source with the same experimental setup.
  
Utilizing the spectrum measurements with a substantiated Monte Carlo 
simulation, we compared the unique quasi-triangular shape induced by the 
691.6 keV electrons and nuclear recoils in the data to the Monte Carlo 
simulations for the two models. Shape analysis was used to verify the 
quasi-triangular shape of spectra in the data and Monte Carlo simulations by 
analyzing data points in the region of interest. We found a good agreement 
between the measurements and the Lindhard (with $k$ = 0.159) and Barker-Mei models. 

In this paper, we corroborate the ionization efficiency models proposed by 
Lindhard {\it et al} and Barker-Mei with a neutron induced $E_0$ transition for 
$^{72}$Ge. The experimental design is discussed in Section 2, followed by data 
analysis in Section 3. The Monte Carlo simulation is demonstrated and validated 
in Section 4, and the comparison with data described in Section 5. Section 6 
summarizes our results.

\section{Experimental Design}
The germanium detector used in our experiment was an old coaxial detector from 
Princeton Gamma Tech, model  RG11B/C~\cite{princeton}. Its linearity of energy 
response between consecutive calibrations was within 0.35\%. However, the full 
width at half maximum (FWHM) was 7.1 keV at the 1173 keV $^{60}$Co peak, a 
factor of two worse in energy resolution than a new germanium detector. 
Nevertheless, the detector was sufficient for measuring neutron induced 
internal conversion. We used an $^{241}$Am-$^{9}$Be (AmBe) source with neutron 
energies ranging from $\mathtt{\sim}$1.0 to 11.2 MeV, at a frequency of 
100 Hz~\cite{ambe}.

An AmBe neutron source produces neutrons in four discrete groups: $n_0$, $n_1$, 
$n_2$, and $n_{3}$, which populate the ground state, the 4.443 MeV level, the 
7.65 MeV level, and the 9.64 MeV level of the $^{12}$C product nucleus, 
respectively~\cite{kwgl, fdeg}. The $n_1$ group neutrons, accompanied by 
4.443 MeV gamma rays, dominate the production of neutrons in an AmBe 
source~\cite{kwgl, fdeg}. This feature allowed us to set up a coincidence 
measurement using sodium-iodide (NaI) detectors (Bicron model number 
3M3/3~\cite{bnai}) with a threshold of above 1 MeV. Utilizing this coincidence 
method, the NaI detectors measured gamma rays while neutrons were detected 
with the germanium detector. This coincidence pattern required that the NaI 
and Ge detectors both trigger in order for an event to be recorded, suppressing 
random background events generated by gamma rays from surrounding materials. We 
took approximately 22 days of data with three NaI detectors and 13 days of data 
with two NaI detectors. 

\begin{figure}[htb!!]
\includegraphics[width = 0.46\textwidth] {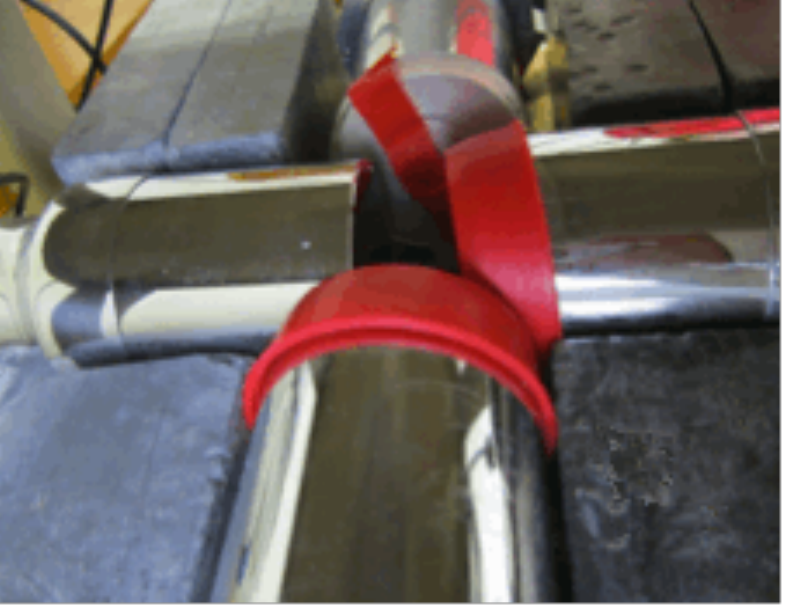}
\includegraphics[width = 0.48\textwidth] {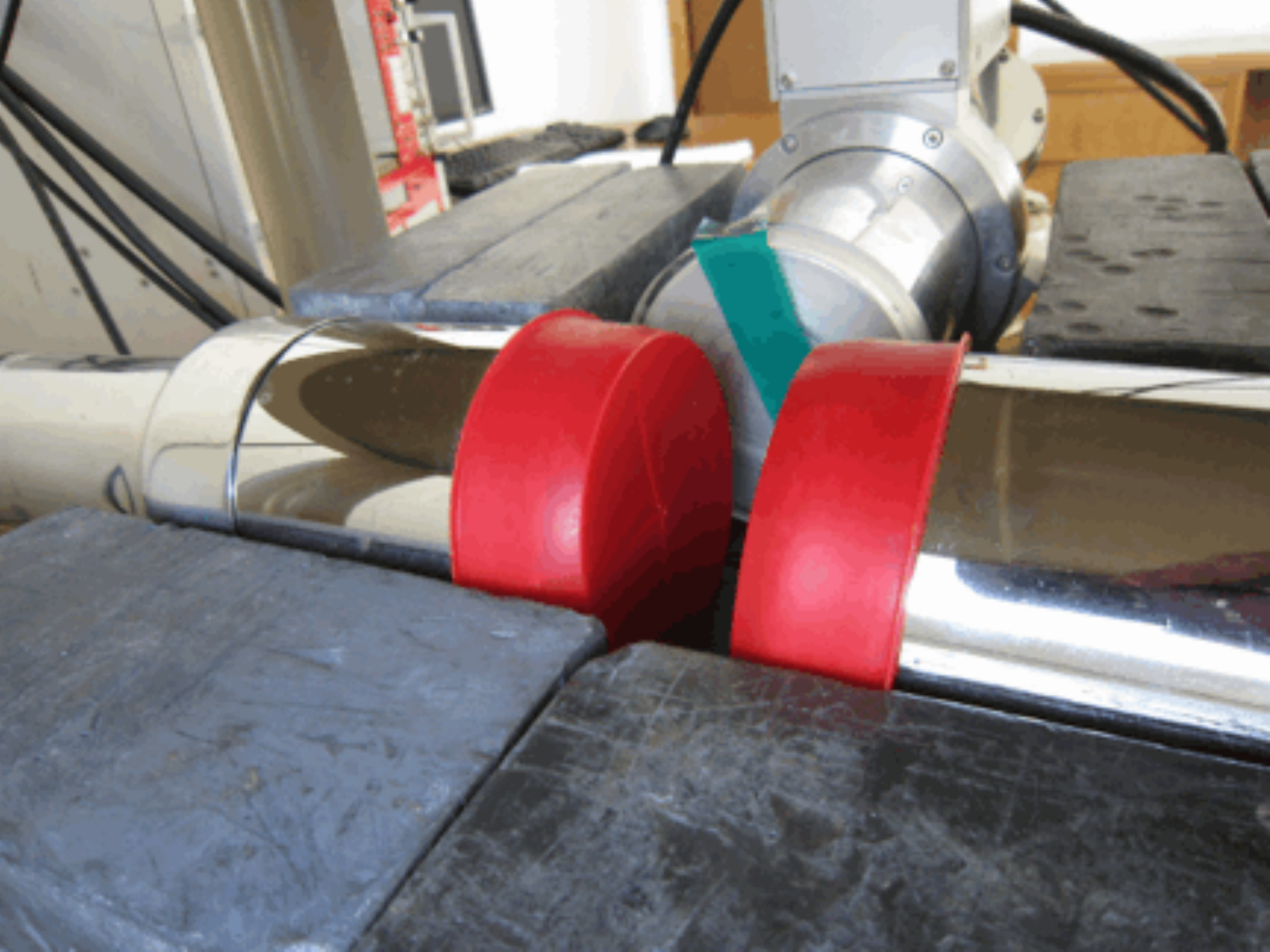}
\caption{\small{Experimental setup for Ge and NaI detectors. Left: Three NaI detectors. Right: Two NaI detectors.}}
\label{fig:ExpSetup}
\end{figure}

\begin{figure}[htb!!]
\includegraphics[angle=0,width=12.cm] {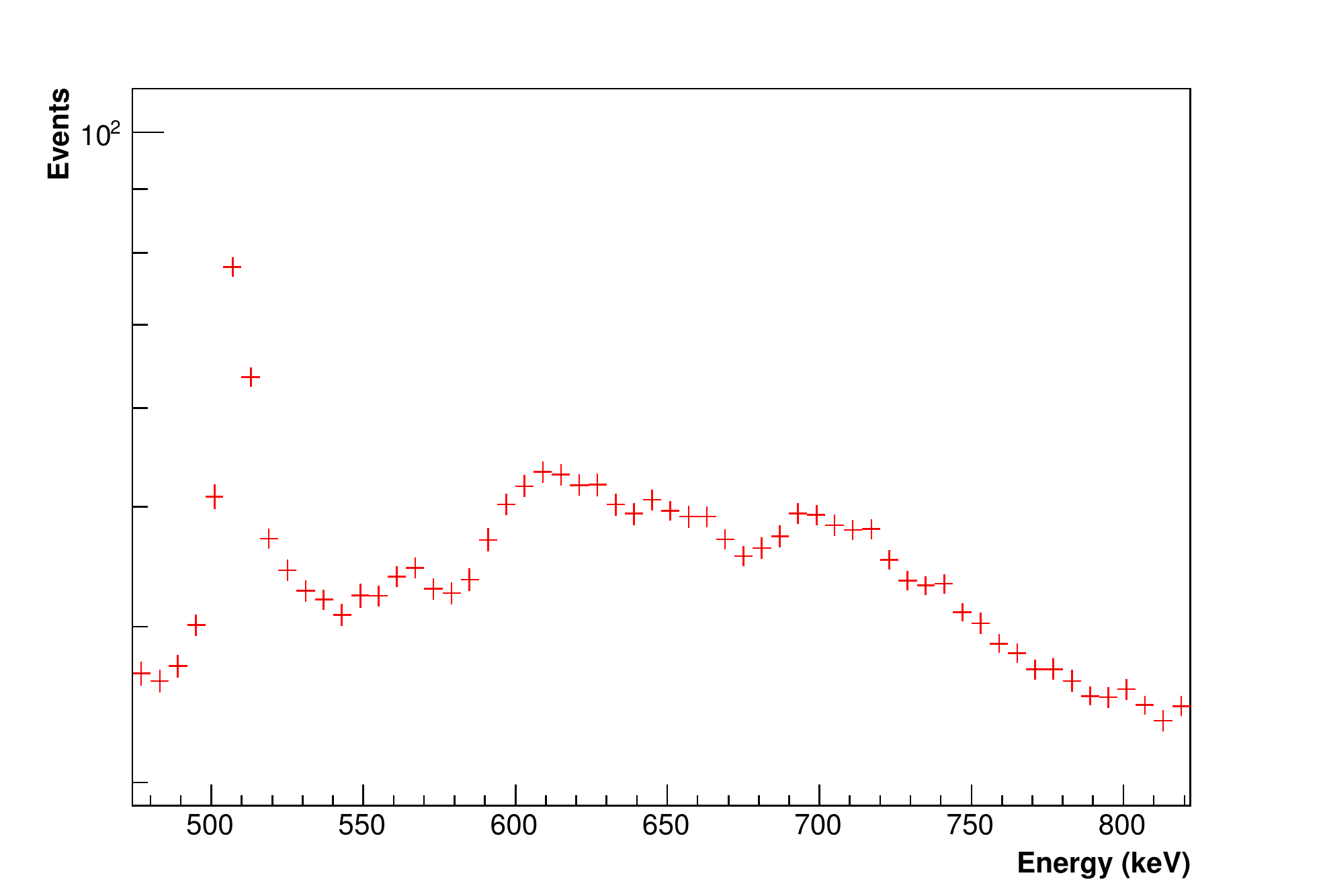}
\caption{\small{Data taken with AmBe source after 35 days.}}
\label{fig:ExpData}
\end{figure}

Due to low neutron emission, the source was placed on the center of the Ge 
detector cap and held in place by electrical tape. When using three NaI 
detectors, they were placed on the right, the left, and directly in front of 
the germanium detector. When using two NaI detectors, they were placed on the 
right and on the left of the germanium detector. Fig.~\ref{fig:ExpSetup} shows 
the experimental set-up. 
The data acquisition was performed using a National Instruments PXI-1031 
system~\cite{xia} and Igor Pro 4.07 software~\cite{igor}. Each run lasted 
approximately 4.5 days. The data from each individual run was added 
consecutively off-line using analysis code from the Root software 
package~\cite{root}. Evidence of the 691.6 keV $E_0$ transition peak can be 
seen in Fig.~\ref{fig:ExpData}.

\section{Experimental Results}
\subsection{Data Analysis}
After 35 days, the number of events in the region of interest (675 
$\mathtt{\sim}$ 765 keV) was approximately 500. This is consistent with the 
known neutron emission from the AmBe source and the coincidence method, which 
suppresses random background events from gamma rays, as verified by the Monte 
Carlo simulation. Because of the small sample size of valid data in the region 
of interest, the bin size in the histogram was set to 6 keV in order to 
mitigate statistical fluctuation. As shown in Fig.~\ref{fig:ExpData}, 
there are several peaks near the region of interest. We identified these peaks 
in order to help understand the processes that took place within our detector 
as well as possible sources of contamination.

The first peak can be fitted using a standard Gaussian distribution and linear 
background distribution. This is given by:
\begin{equation}
p_0\cdot exp\left[-\frac{1}{2}\left(\frac{E-p_1}{p_2}\right)^{2}\right]-p_3 +p_4E,
\end{equation}
where $p_0$ = 37 $\pm$ 1.4 is the normalization constant, $p_1$ = 508 $\pm$ 0.2 
keV is the center value, $p_2$ = 4.7 $\pm$ 0.2 keV is the Gaussian width, 
$p_3$ = -65 $\pm$ 12 is a constant, $p_4$ = 0.2 $\pm$ 0.02 is the slope, and 
$E$ is the energy in keV. The peak at 511 keV is mainly from the annihilation 
of e$^{-}$e$^{+}$ pairs induced by cosmic rays passing through the surrounding 
materials with other minor contributions. Most notable is the 4.443 MeV induced 
e$^-$e$^+$ in the surrounding materials. The positrons can annihilate with electrons 
producing 511 keV gamma rays, which enter the germanium detector. 
The fitted central value of 508 keV is slightly lower than the expected 511 keV, 
but still within the margin of error for the given bin size. 

The remaining peaks can be fitted using a Moyal distribution and linear 
background. The Moyal distribution used is~\cite{moyal}:
\begin{equation}
\Psi = \sqrt[]{\frac{1}{2\pi}exp\left[-\left(R\left(E-E_{mpv}\right)+exp\left[R\left(E-E_{mpv}\right)\right]\right)\right]},
\label{moyalFit}
\end{equation}
where $R$ is a constant and $E_{mpv}$ is the most probable value of energy 
deposition in the detector. $R$ and $E_{mpv}$ are physically significant 
parameters. The value of $R$ is related to the reciprocal of the stopping power 
and can be used to calculate average path length for particles in the medium. 
The most probable value of energy, $E_{mpv}$, is the most common energy 
deposition in that region. These parameters are further discussed in Section 3.3 and used 
to interpret the experimental data.

\begin{figure}[htb!!]
\includegraphics[angle=0,width=12.cm] {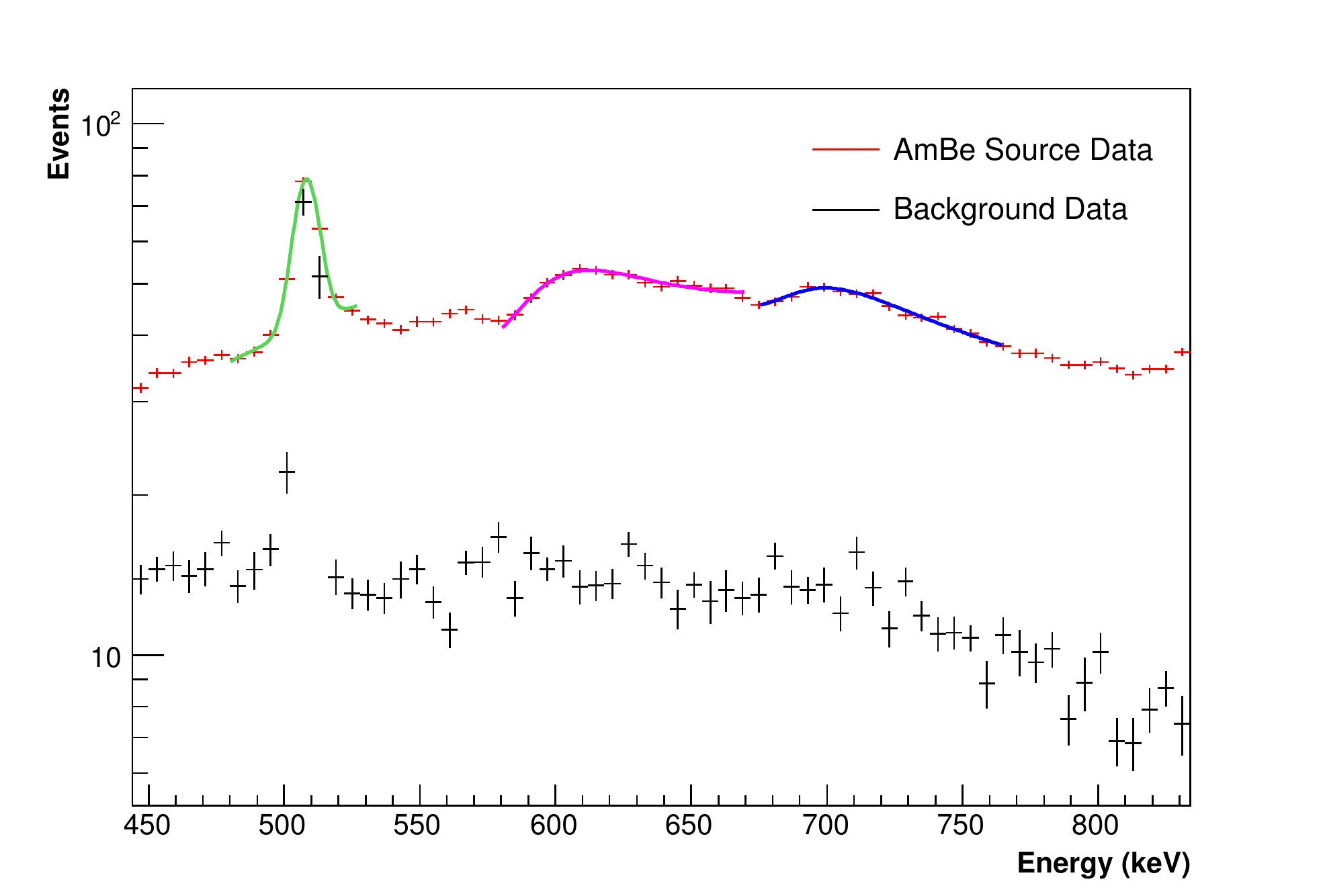}
\caption{\small{Region of interest with all peaks fitted, and the coincidence background spectrum.}}
\label{fig:ExpDataFitAll}
\end{figure}

The peak around 560 keV, is likely caused by events from 
$^{76}$Ge$(n,n'\gamma)$ inelastic scattering. Because of the small number 
of events, the fitting function was only partially accurate. The second peak at 
around 600 keV has the fitted parameters, $R$ = 0.06 $\pm$ 0.02 and 
$E_{mpv}$ = 609 $\pm$ 3.6 keV. This peak is likely the combination of inelastic 
scattering from $^{74}$Ge$(n,n'\gamma)$ and the neutron capture on $^{73}$Ge, 
$^{73}$Ge$(n,\gamma)$. Finally, the peak of interest, 691.6 keV, exhibits the 
internal conversion of $^{72}$Ge$(n,n'e)$. The fitted parameters are, 
$R$ = 0.06 $\pm$ 0.02 and $E_{mpv}$ = 705 $\pm$ 3.0 keV.
All fits in the region of interest are shown in Fig.~\ref{fig:ExpDataFitAll}.

We also took a background spectrum using the coincidence method to identify any 
random coincidence. This spectrum is also shown in Fig.~\ref{fig:ExpDataFitAll}.
The background data was taken using two sodium iodide detectors (see right 
of Fig.~\ref{fig:ExpSetup}) for 4.6 days. 

\begin{figure}[htb!!]
\includegraphics[angle=0,width=12.cm] {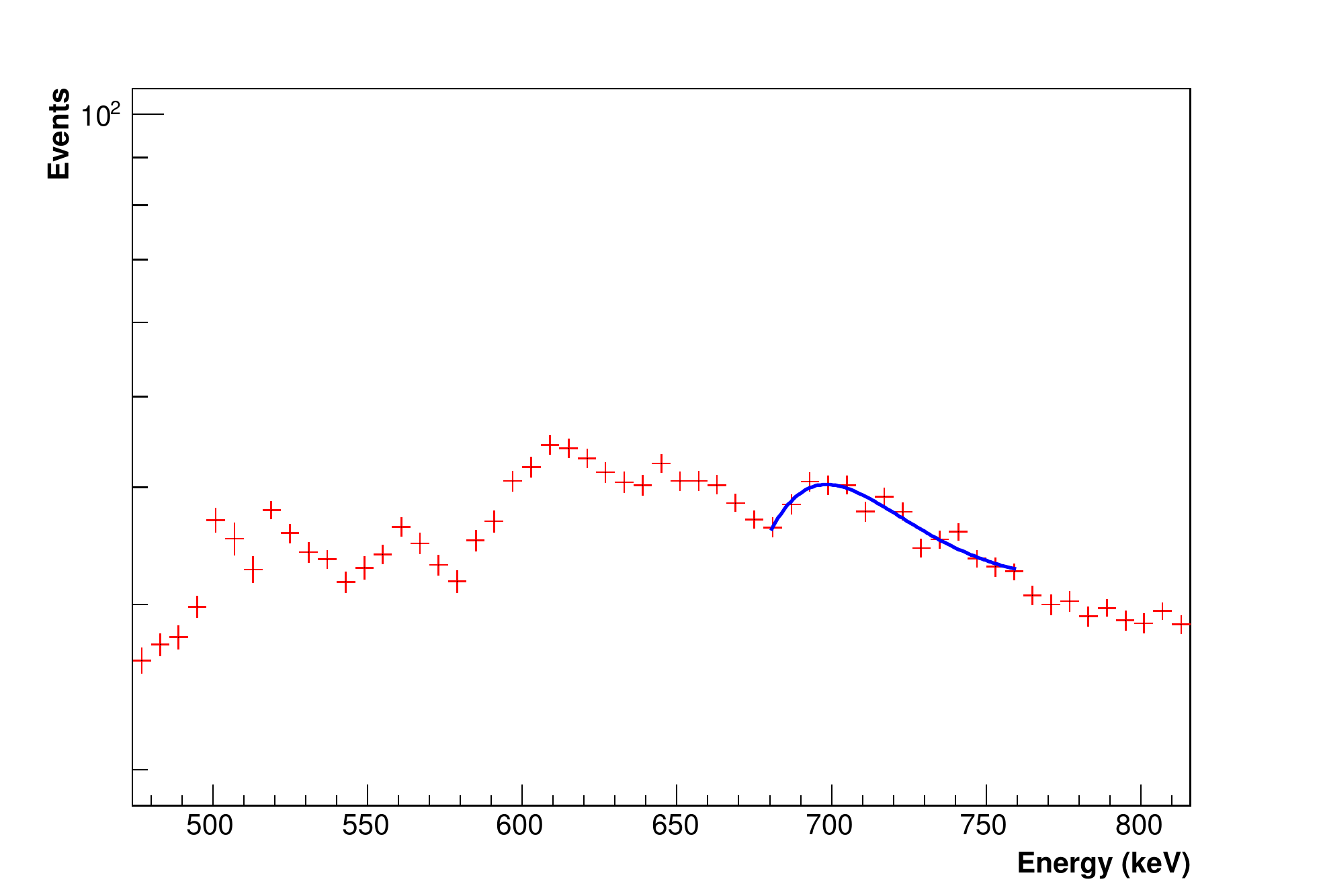}
\caption{\small{Region of interest with the background subtracted from the original data files.}}
\label{fig:BkrdSubtracted}
\end{figure}

After the background spectrum was collected, we subtracted it from the AmBe 
data sets as shown by Fig.~\ref{fig:BkrdSubtracted}.
The 511 keV peak was significantly reduced, verifying our conjecture on the 
main origin of this peak as induced pair production from cosmic rays in the 
surrounding materials. The subtracted data set was also fitted using a Moyal 
distribution and linear background. It was found that the most probable value 
of energy deposition in the detector for the $E_0$ transition 
was 696 $\pm$ 1.4 keV. Thus the percent difference of the AmBe data with and 
without background subtraction is 1.3\%.

\subsection{Error Analysis}
We have identified the following sources of systematic error for our data 
analysis: the energy scale and the energy resolution. The associated energy 
scale error is 3 keV, which is from the fit of the known 511 keV peak. The center value of the fitted 
function was 508 keV which is a difference of 3 
keV from the known value of the peak. This shift in energy is from the energy
calibration used to convert to the energy scale. As stated previously, there 
was a 0.35\% error in successive energy calibrations. Using a more accurate 
energy scale to produce a peak at 511 keV caused other
energy regions to become less accurate. For energy resolution, the 
value from the 662 keV peak of $^{137}$Cs was used, it is nearest of the calibration 
sources to our region of interest. The energy resolution for this peak was 
3.45 keV. Adding two errors in quadrature, since they are independent of each 
other, causing the resulting systematic error to be 4.6 keV. Thus, the larger bin 
size of 6 keV was used to accommodate for this error.

Statistically, there are approximately 500 events in the region of interest 
(675 keV $\mathtt{\sim}$ 765 keV) which gives a statistical error of 4.5\%.
For data with the background subtracted out, the error associated with the most 
probable value is 696 $\pm$ 4.6 (sys) $\pm$ 1.4 (stat).

\subsection{Interpretation of Results}
Electrons and nuclear recoils that travel through germanium with a high 
momentum lose energy by exciting and ionizing the germanium atoms. The average 
amount of energy lost can be calculated with the Bethe-Bloch 
equation~\cite{bbe}. However, the energy transfer is not a continuous process. 
It occurs through random collisions during which various amounts of energy can 
be transferred. The energy loss in the detector can be described by the 
Moyal distribution, Eq.~\ref{moyalFit}, as shown in 
Fig.~\ref{fig:ExpDataFitAll}.

As shown in Eq.\ref{moyalFit}, $R$ is the reciprocal of the density, $\rho$, times the average 
path-length of charged particles in  the detector, $d$, times the parameter, 
$K$, as given by the Bethe-Block formula:
\begin{equation}
R = \frac{1}{K \cdot \rho \cdot d},
\end{equation}
where $K$ = 4$\pi N_a m_e c^2 r_e^2 z^2 \frac{Z}{A} \frac{1}{\beta^2}$ is 
related to the stopping power, with constants $N_{a}$, Avogadro's number; 
$m_{e}$, the mass of electrons; $c$, the speed of light; $r_e$, the Bohr 
radius; $z$, the electron charge; $Z$, atomic number of target; $A$, the atomic 
mass number of target; and $\beta$, the speed of charged particles divided by 
$c$. The density of germanium is $\rho$ = 5.323 $g/cm^3$. 

Since electronic stopping power is different from nuclear stopping power, 
the value of $R$ is very different for pure electronic recoils than nuclear 
recoils. Thus, the value of $R$ can be used to derive the average path-length 
for a given electronic recoil or nuclear recoil. However, the value of $R$ from 
our measurements is a combination of electronic and nuclear recoils, and it 
cannot be used to directly determine the average path length for either. 
Nevertheless, we can use the most probable value of energy deposition, which is 
related to the stopping power multiplied by the average path length, to 
determine the average path length.

Given $E_{mpv}$ = 696 $\pm$ 4.6 (sys) $\pm$ 1.4 (stat) keV, obtained from the 
fitted function in Fig.~\ref{fig:BkrdSubtracted}, is a convolution of the 
691.6 keV energy deposition and the nuclear recoils, then we contend that 
696 keV = $\frac{dE}{dX}$ $\cdot d\cdot \rho$, where $\frac{dE}{dX}$ is the mass 
stopping power in keV cm${^2}$/g. Since there are both electronic and nuclear 
recoils, we can rewrite this as:
\begin{equation}
696\pm4.6\pm1.4\hspace{0.1in}keV = \left(\frac{dE}{dX_e} \cdot d_e + \frac{dE}{dX_n} \cdot d_n\right) \cdot \rho,
\end{equation}
where $\frac{dE}{dX_e}$ = 1301 keV cm$^{2}$/g~\cite{nist} and $\frac{dE}{dX_e} \cdot d_e \cdot \rho$ = 691.6 keV; thus $d_e$ = 691.6 /($\frac{dE}{dX_e} \cdot \rho$) = 0.1 cm, which is the average path length in germanium
for electrons with an energy of 691.6 keV. 
Therefore, we have
\begin{equation}
696\pm4.7\pm1.4\hspace{0.1in}keV = 691.6\hspace{0.1in}keV  + \frac{dE}{dX_n} \cdot d_n \cdot \rho
\end{equation}
and  
\begin{equation}
4.4\pm0.007\pm0.002\hspace{0.1in}keV = \frac{dE}{dX_n} \cdot d_n \cdot \rho.
\label{eq:recoil}
\end{equation}
From Eq.\ref{eq:recoil}, we can conclude that the most probable nuclear 
recoils that we measured in the detector have an average of 4.4 $\pm$ 0.007 
(sys) $\pm$ 0.002 (stat) keV electronic equivalent energy, which corresponds to 
approximately 17.5 $\pm$ 0.12 (sys) $\pm$ 0.035 (stat) keV nuclear recoil 
energy from ionization efficiency in the Barker-Mei model.

To determine $d_n$, we can use $\frac{dE}{dX_n}$ = 2341720 keV cm$^{2}$/g for a 
nuclear recoil of 17.5 keV~\cite{dbar}. Thus, 
$d_n$ = 17.5 keV /($\frac{dE}{dX_n} \cdot \rho$) = 0.014 $\mu$m is the average 
path length in germanium for 17.5 keV nuclear recoils. This determines the 
range of low energy nuclear recoils in a germanium detector.

\section{Monte Carlo Simulation of $E_0$}
An accurate Monte Carlo simulation is needed to determine the validity of the 
Lindhard and Barker-Mei models when compared to collected data.
Two crucial steps were taken before creating a Monte Carlo simulation that would
determine nuclear recoil energy: 1) the $E_0$ transition in 
Geant4.9.2~\cite{geant4} was fixed and 2) the Monte Carlo simulation was 
verified with a well-known gamma-ray source.

\subsection{Fixing the $E_0$ transition in Geant4.9.2}
By studying the inelastic scattering processes that contribute to 
internal conversion, we found that the $E_0$ transition code is included 
in Geant4, but is missing neutron data for $^{72}$Ge. Specifically, 
Geant4 does not provide $\gamma$/$e^−$ transition data in each energy level 
(Data Type = 12) and has no cross section data corresponding to the $E_0$ 
transition. 

\begin{figure}[tb!!]
\includegraphics[angle=0,width=12.cm] {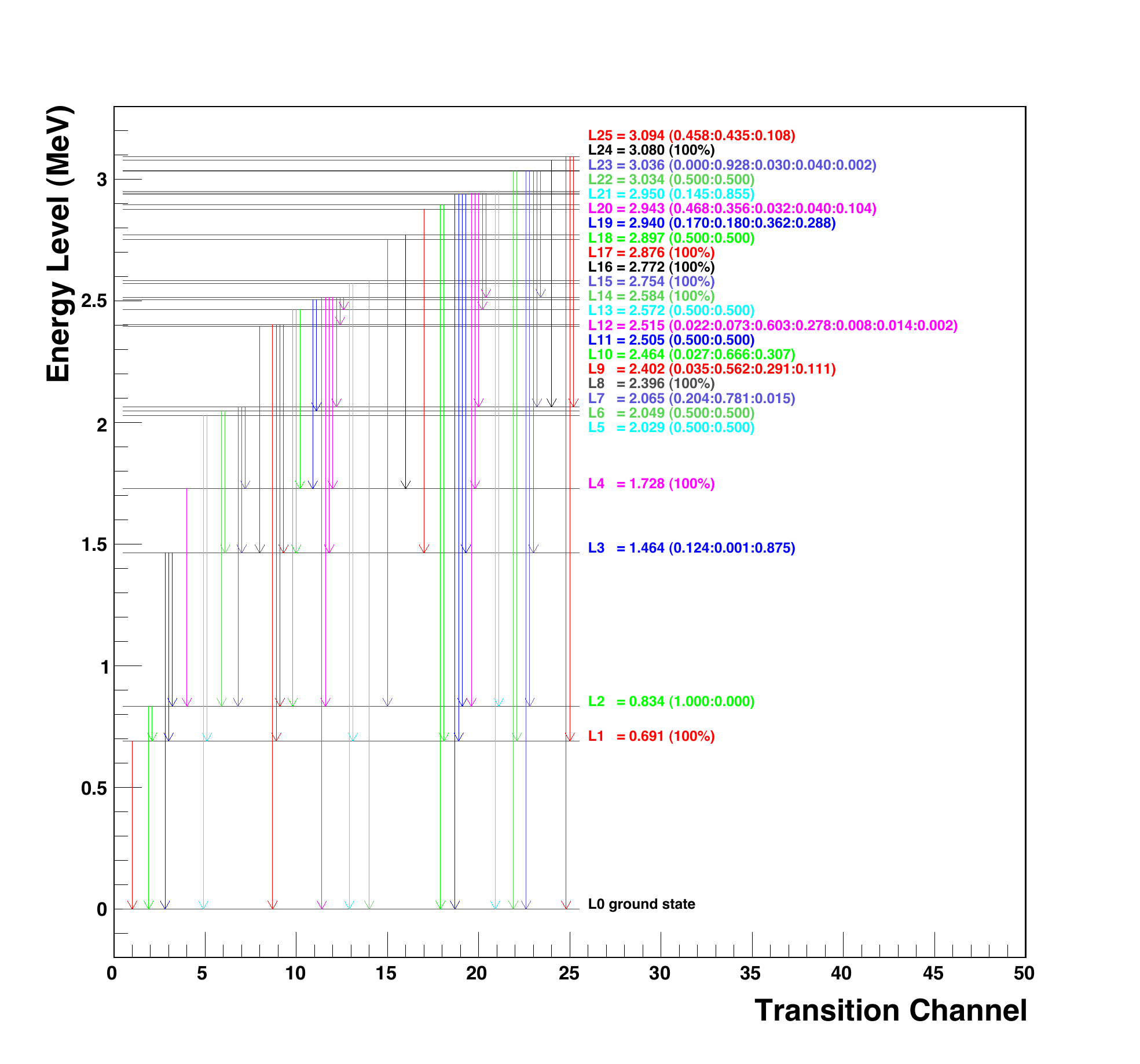}
\caption{\small{The orbital $e^-$ transition ratio of $^{72}$Ge in its excited states.}}
\label{fig:transmap}
\end{figure}

In order to make the internal conversion process available in our Geant4 
simulation, we created a $\gamma$/$e^−$ ratio in the database for $^{72}$Ge 
with Geant4.9.2 several years ago 
(see our correction in Ref.~\cite{prob4}). In addition, we calculated all the
cross-sections of the $^{72}$Ge$(n, n')$ reaction using TALYS~\cite{ajk}, 
a reliable software for the simulation of nuclear reactions. We then
converted the transition cross-section data into the transition ratio in Geant4.9.2 as shown 
in Fig.~\ref{fig:transmap} (we utilize Geant4.9.2 for this work because the
E$_0$ transition problem was amended several years ago).

The independence of this ratio can be cross-checked by looking at the reactions 
in different energy scales. Eventually, the TALYS data was converted to 
Geant4 format where it could be used in the internal conversion process (see 
the supplemental data used in Geant4.9.2 from Ref.~\cite{prob4}).

During the $E_0$ transition, a characteristic X-ray, or Auger electron, is 
produced simultaneously with the conversion electron. This process was not 
included in the Geant4 code. However, since this is a complicated 
process that involves binding energy, it would require editing for all the
elements in the Geant4 package. For the sake of simplicity, we combined 
the X-ray and conversion electron together, since they both contribute to the 
total energy deposition (in a solid germanium detector only). A more general 
solution will be required when considering other materials.

A simulation was created from the Geant4.9.2 and G4NDL3.13 packages that 
includes the additional transition ratio data. For further information, readers 
can refer to Ref.~\cite{dmm2} for detailed geometry and experimental 
framework. After the missing data was added, the $E_0$ transition of 
$^{72}$Ge$(n, n'e)$ could be simulated using Geant4.9.2. This makes the 
Geant4.9.2 simulation more reliable for dark matter searches utilizing 
germanium detectors.

\subsection{Verification of the Monte Carlo Simulation}
\begin{figure}[tb!!]
\includegraphics[width = 0.57\textwidth] {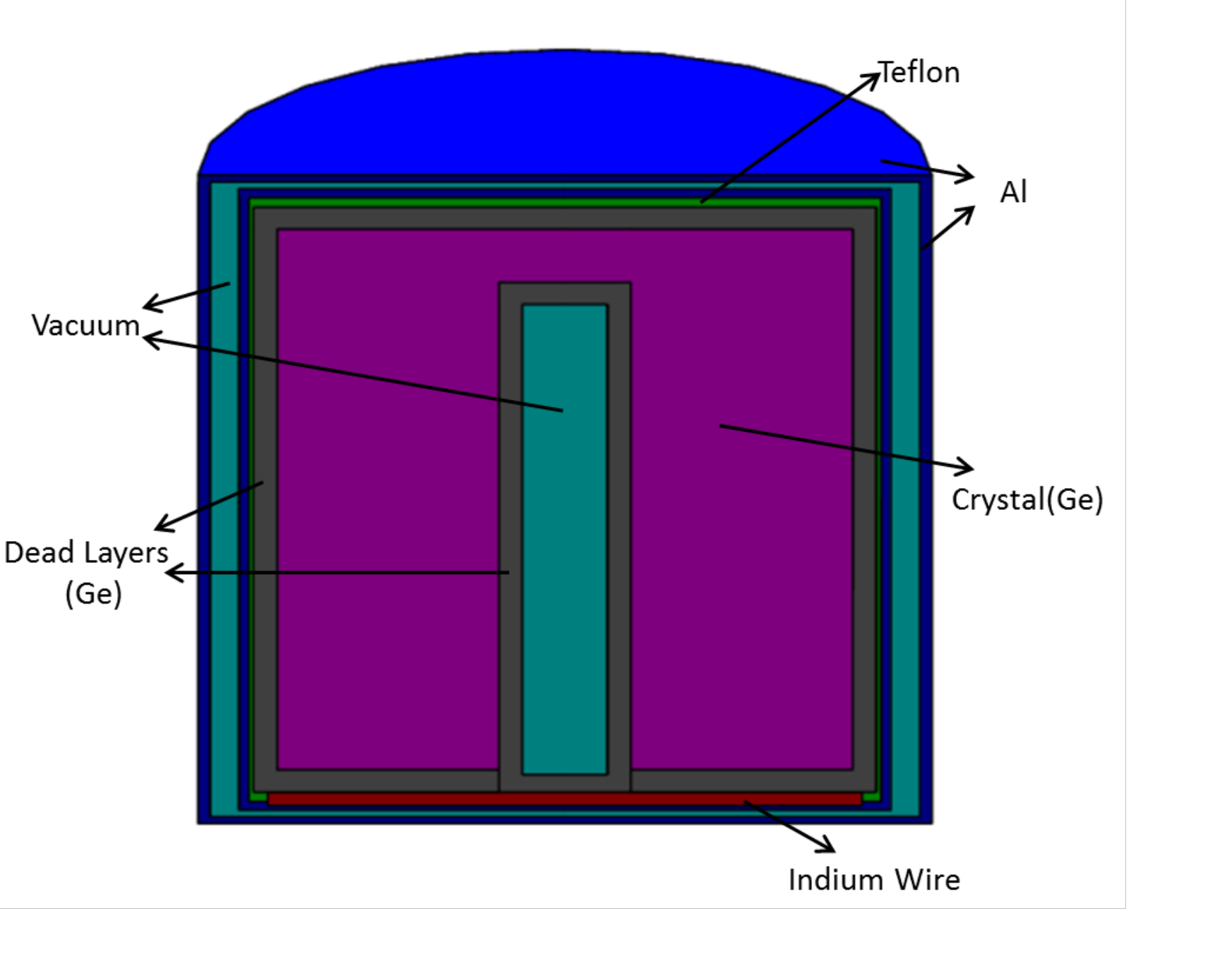}
\includegraphics[width = 0.45\textwidth] {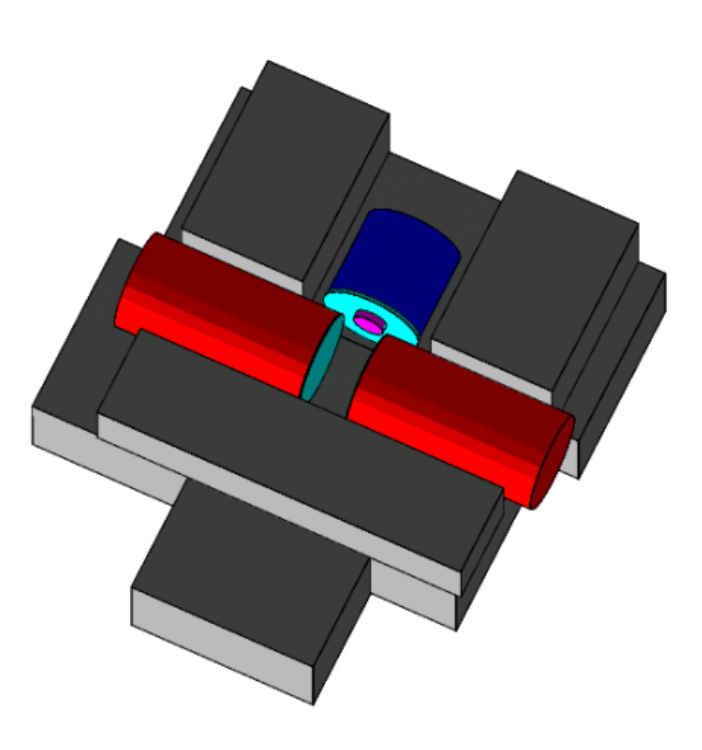}
\caption{\small{Simulated geometry for Ge and NaI detectors. Left: Cross section of Ge detector. Right: Ge detector (Blue), two NaI detectors (Red) and Lead bricks (Gray).}}
\label{fig:GeGeometry}
\end{figure}

\begin{figure}
\includegraphics[angle=0,width=12.cm] {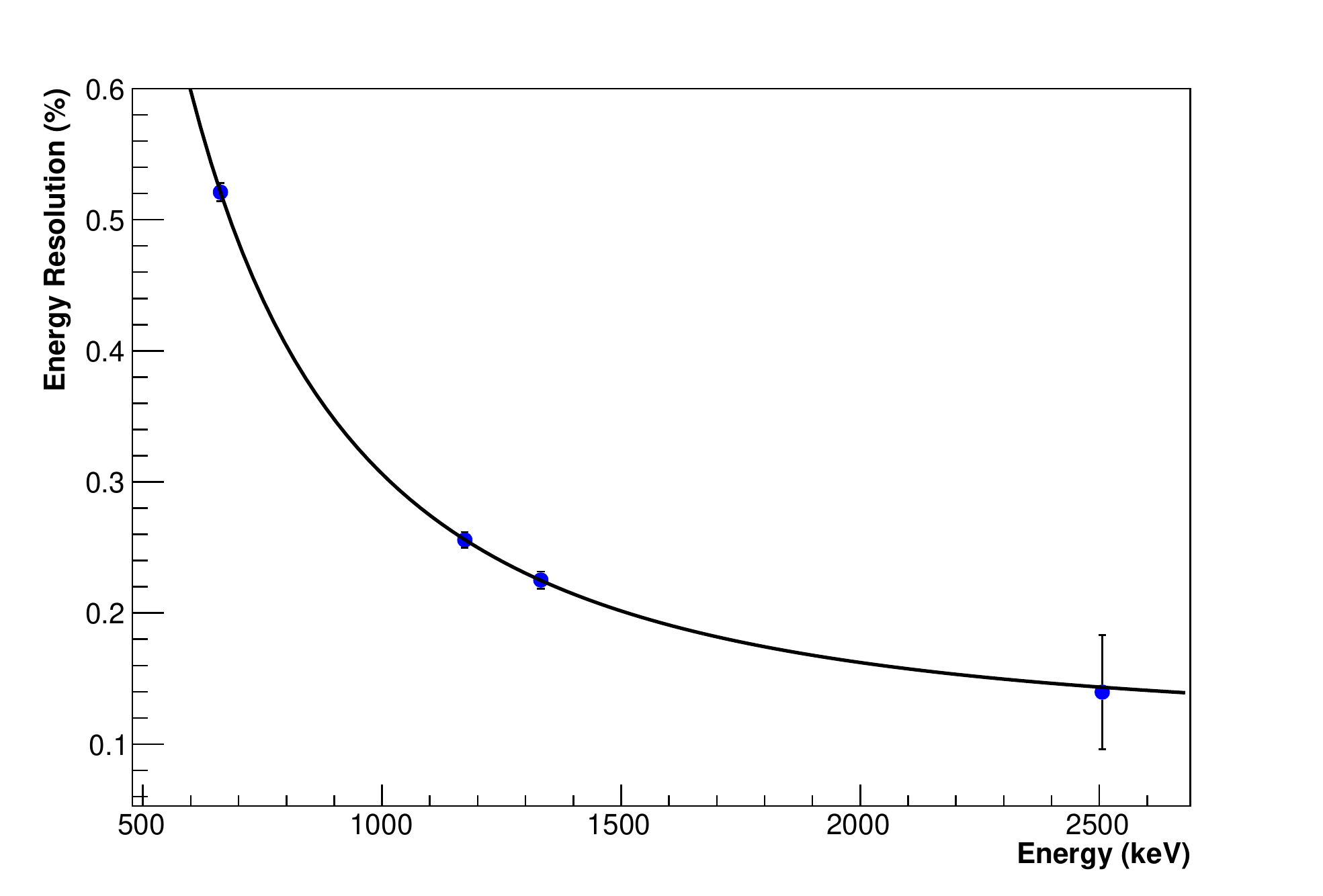}
\caption{\small{Percentage of relative energy resolution as a function of
energy. The error bars are statistical errors. }}
\label{fig:EnRes}
\end{figure}

\begin{figure}[tb!!]
\includegraphics[angle=0,width=12.cm] {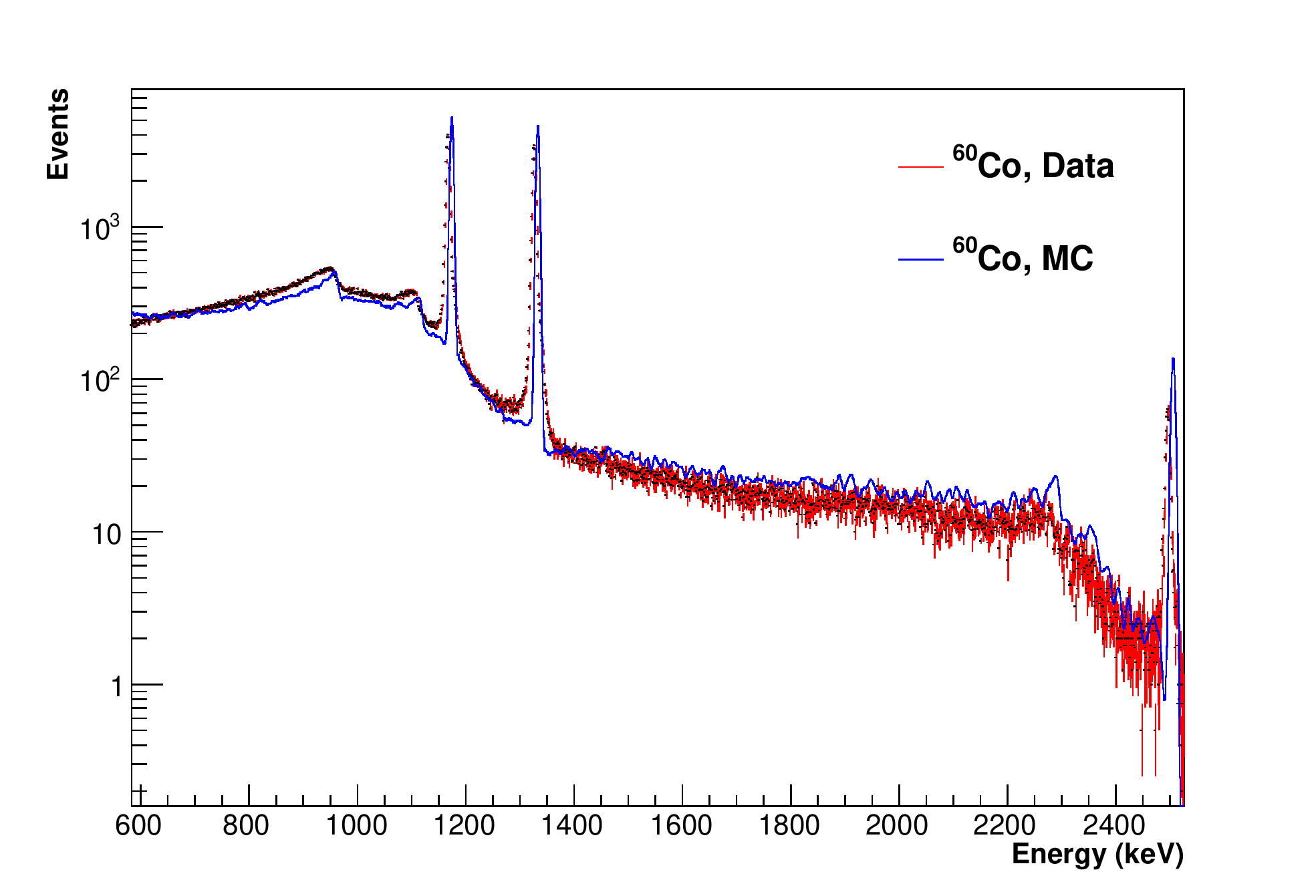}
\caption{\small{Comparison between MC and data in energy deposition spectrum.}}
\label{fig:Co60}
\end{figure}

To obtain reliable results from our Monte Carlo, it was essential to verify 
the simulation. $^{60}$Co, with an original radioactivity of 1.0 $\mu$Ci, was 
used to validate the simulation. The $^{60}$Co source was mounted on the center 
of the Ge detector cap as shown in Fig.~\ref{fig:ExpSetup} (right).
The corresponding geometry used in the Monte Carlo simulation is shown in 
Fig.~\ref{fig:GeGeometry}.
Since the germanium detector is aged, the Monte Carlo geometry was modified to 
include a dead layer of thickness 2.5 mm (see Fig.~\ref{fig:GeGeometry}, left). 
We also implemented a smearing process by applying energy resolution to the 
spectrum of deposited energy in the active germanium. When fitting the energy 
resolution, these four peaks were identified: 1173 keV, 1332 keV, 2506 keV and 
662 keV. The first three peaks are gamma rays from the $^{60}$Co source and the 
last peak is a gamma ray from $^{137}$Cs source. The best fit function 
is presented in Eq.~\ref{EnResFit}:
\begin{equation}
0.2017\pm0.00846\times\sqrt{0.3+(0.001E)^{(-1.96\pm1.69)}+(0.001E)^{(-3.436\pm1.262)}},
\label{EnResFit}
\end{equation}
where $E$ is the energy in keV. Under the square root, the first term, 0.3, is 
the percentage of energy resolution, for a new Ge detector; the second and 
third terms are due to noise and the age of the Ge detector. The percentage of 
relative energy resolution as a function of energy is plotted in 
Fig.~\ref{fig:EnRes}
Using this fitted function, we applied the energy resolution to the range 600 
keV $\mathtt{\sim}$ 2510 keV and obtained agreement between the 
experimental data and Monte Carlo simulation as shown in Fig.~\ref{fig:Co60}. 
This validates that our Monte Carlo can be reliably used for gamma-ray 
simulation.

\section{Monte Carlo Simulation of Nuclear Recoils}
Using the modified Geant4.9.2 package, we performed simulations to verify the 
accuracy of the Lindhard and Barker-Mei models with an AmBe neutron source. The 
experimental setup for our Ge and NaI detectors, as shown in 
Fig.~\ref{fig:ExpSetup}, was simulated according to the dimension and material 
information provided by the manufacturer~\cite{manufacturer}. Since the 
$^{72}$Ge$(n,n'e)$ reaction causes a quasi-triangular shape in the data, we 
expect to see this same feature in our Monte Carlo simulation if the Lindhard 
and Barker-Mei models are accurate. In order to provide an accurate simulation,
the geometry, as well as the AmBe source generator, need to be implemented 
correctly.

\subsection{AmBe Neutron Generator}
Because the AmBe source was placed very close to the germanium detector, gamma 
rays emitted from the source have a greater chance of entering the germanium 
detector and causing contamination in the region of interest (675 keV 
$\mathtt{\sim}$ 765 keV). Thus, in the simulation it is necessary to account 
for all potential gamma rays emitted from the AmBe source.

Two reactions, shown in  Eqs.~\ref{am241} and \ref{be9}, occur in the AmBe
source:
\begin{equation}
^{241}Am \rightarrow ^{237}Np + \alpha + \gamma
\label{am241}
\end{equation}
\begin{equation}
\alpha + ^{9}Be \rightarrow ^{12}C + n + \gamma.
\label{be9}
\end{equation}

Eq.~\ref{am241} shows the decay of $^{241}$Am to $^{237}$Np, which causes the 
emission of alpha particles and gamma rays. Eq.~\ref{be9} shows the reaction 
between an alpha particle and $^{9}$Be. In Eq.~\ref{be9}, the energy of gamma 
rays emitted depends on the resulting state of $^{12}$C, which is 4.443 MeV 
(for the 1st excited state), 7.65 MeV (for the 2nd excited state) and 9.64 MeV 
(for the third excited state). Using the gamma ray energy and recoil energy of 
$^{12}$C, the resulting neutron energy can be calculated by applying energy 
conservation to Eq.~\ref{be9}.

Fully absorbed gamma rays (from Eq.~\ref{be9}) will not be in the region of 
interest (675 keV $\mathtt{\sim}$ 765 keV) since their energies are at a few 
MeV. However, the Compton continuum of their interaction can contribute to the 
region of interest. In addition, gamma rays (from Eq.~\ref{am241}) with 
energies 26.34 keV (a branching ratio of 2.4\%), 59.54 keV (a branching ratio 
of 35.9\%) and 722.01 keV (a branching ratio of 0.000196\%)~\cite{web} 
can contribute to the region of interest by occurring in coincidence with the 
E$_0$ transition (26.34 keV and 59.54 keV) or by becoming fully absorbed 
(722.01 keV). We generated these three gamma rays in our Monte Carlo simulation 
along with gamma rays at energies of 4.443 MeV, 7.65 MeV, and 9.64 MeV caused 
by the transitions in the excited $^{12}$C nucleus.

\begin{figure}[tb!!]
\includegraphics[angle=0,width=12.cm] {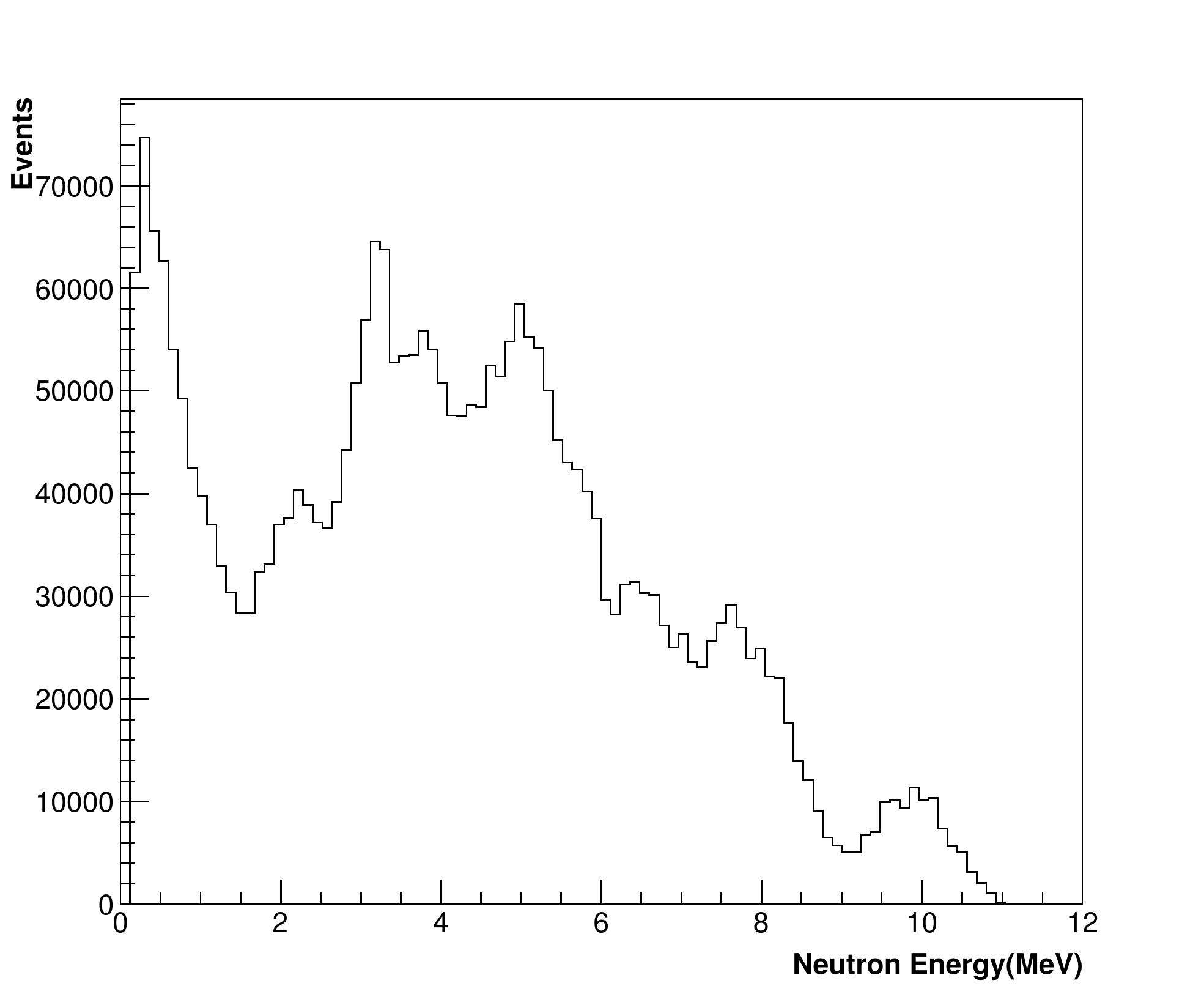}
\caption{\small{Simulated neutron energy spectrum from AmBe neutron source.}}
\label{fig:AmBeNeutron}
\end{figure}

The simulated neutron energy spectrum from the AmBe neutron source is shown in 
Fig.~\ref{fig:AmBeNeutron}. The spectrum agrees with the prediction from 
Marsh et. al.~\cite{jwm}.
After this validation, the AmBe source generator was implemented in the Monte 
Carlo simulation.

\subsection{Verification of the $E_0$ Transition in Simulation}
The simulated results are presented as a spectrum of energy deposited in the 
germanium detector after the application of smearing, which is the process of 
accounting for energy resolution in a germanium detector.

\begin{figure}[tb!!]
\includegraphics[angle=0,width=12.cm] {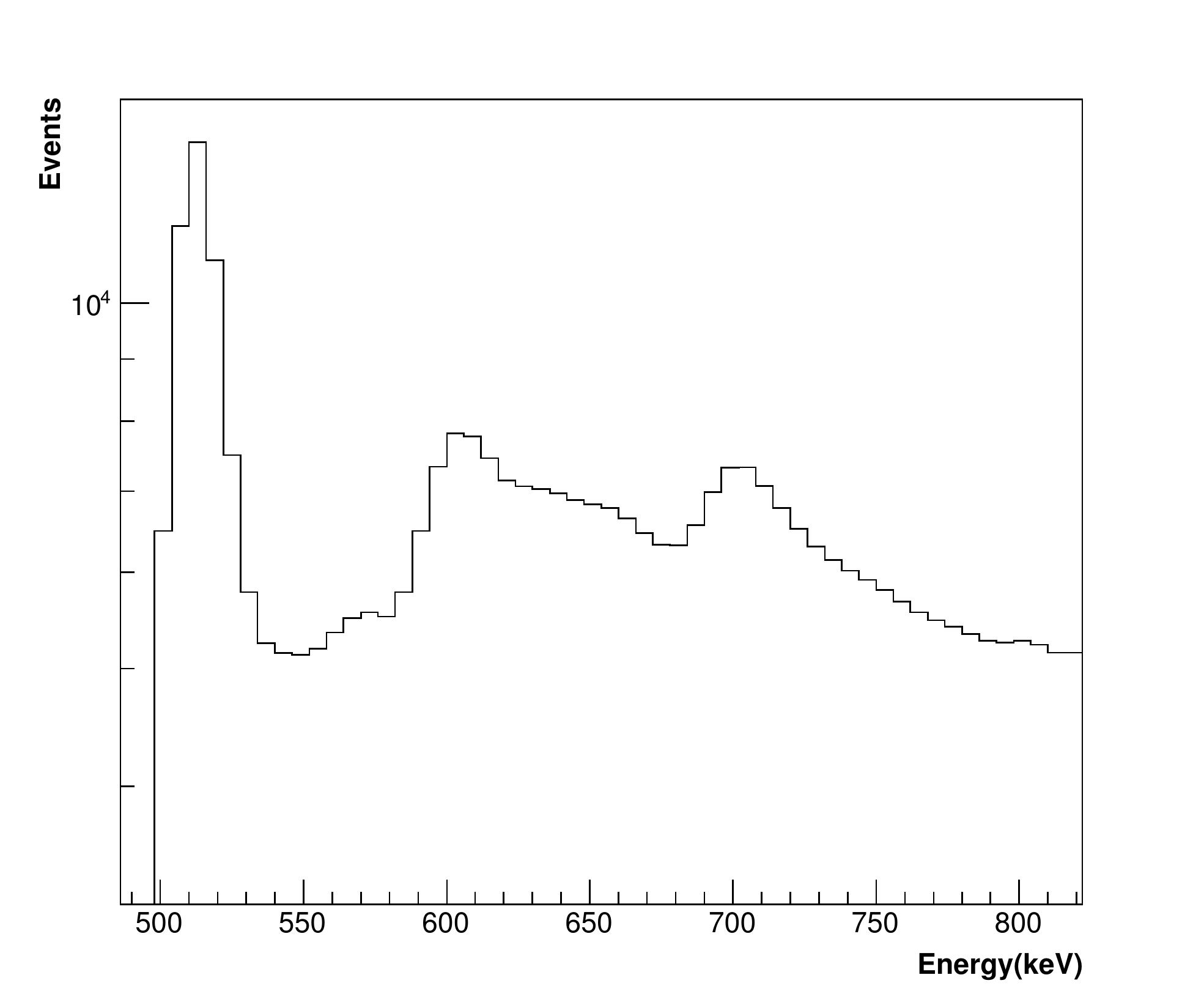}
\caption{\small{The spectrum of energy deposition in the germanium detector after smearing.}}
\label{fig:depHPGeAfter}
\end{figure}

Gaussian and Moyal distributions with fitted parameters from our data 
(Section 3) have been incorporated into the smearing of different peaks in the 
region of interest. We used the model proposed by 
Lindhard {\it et al.}~\cite{lind} ($k$ = 0.159) to determine the ionization 
efficiency for a germanium detector. Fig.~\ref{fig:depHPGeAfter} shows the 
simulated energy deposition spectrum after smearing. The $E_0$, 691.6 keV, 
transition is visible, as shown in Fig.~\ref{fig:depHPGeAfter}. Appearance of 
this peak indicates that the internal conversion process has been successfully 
implemented in our Geant4.9.2.
 
\section{Monte Carlo Shape Analysis}
\begin{figure}[tb!!]
\includegraphics[angle=0,width=12.cm] {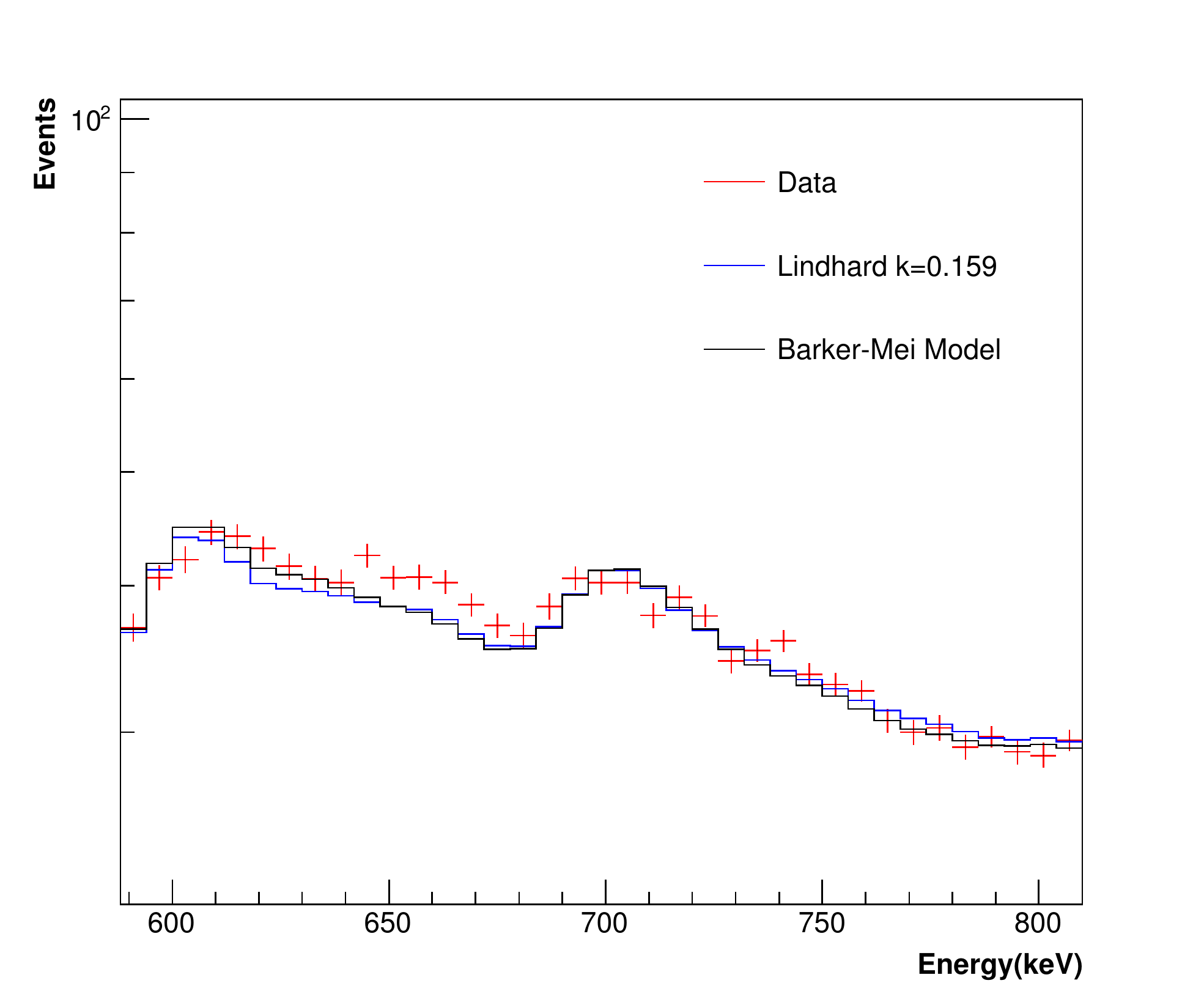}
\caption{\small{Comparison of our Monte Carlo simulation utilizing the Lindhard ($k$ = 0.159) and 
Barker-Mei models and our collected data after background subtraction. The Monte Carlo simulation is normalized to the experimental live-time.}}
\label{fig:aftBkSub}
\end{figure}

After successfully simulating the AmBe neutrons in our Geant4.9.2, we collected 
the energy deposited in the germanium detector. This collected energy spectrum 
was compared with our data after the application of smearing 
(Eq.~\ref{EnResFit}). Two models utilizing ionization efficiency functions were 
applied to the Monte Carlo spectrum: Lindhard $k$ = 0.159 and the Barker-Mei 
model. Normalization was applied to the energy range 675 to 765 keV. By 
overlaying the Monte Carlo energy spectrum and the collected data, we were 
able to perform a shape analysis on the characteristic $E_0$ transition. 
This is demonstrated in Fig.~\ref{fig:aftBkSub}.

The shape analysis was performed on a bin-to-bin basis by comparing the data 
and two Monte Carlo simulations, which correspond to the two models.
The difference in shape between the data and the two Monte Carlo 
simulations was calculated using $\frac{(data - MC)}{(data + MC)/2}$.
Table~\ref{tab:table1} shows the results of this comparison in which the 
largest difference is shown to be less than 4\%.

\begin{table}[tb!!]
\caption{The percent difference between the collected data and the Monte Carlo simulations with two models. 
 }
\label{tab:table1}
\begin{tabular}{|c|c|c|c|c|c|c|c|c|c|}
\hline \hline
 Energy (keV) &691.6& 696 & 702 & 708 & 714 & 720 & 726  &732  & 738\\
\hline 
 Lindhard (k = 0.159) (\%)&2.8& 1.4 & 3.9& 3.1 & 1.0& 0.8 & 2.1& 2.8& 3.0\\
\hline
 Barker-Mei (\%)& 2.6&1.0 & 3.2 &2.3& 0.6& 0.9& 1.8 & 2.2& 2.1\\
\hline \hline
\end{tabular}
\end{table}

Unfortunately, we were not able to include the neutron capture lines of some 
specific even nuclei, such as $^{70}$Ge, in our Geant4.9.2 package due to the 
lack of adequate cross-sections. Thus, at 708.2 keV and 747.7 keV, there is a 
discrepancy between the collected data and Monte Carlo simulations due to the 
neutron capture of $^{70}$Ge$(n,\gamma)$~\cite{dmm2}. There is also 
inconsistency in the energy range 650 $\mathtt{\sim}$ 680 keV that is likely 
due to $^{115}$In$(n,\gamma)$ and $^{206}$Pb$(n,n'\gamma)$~\cite{dmm2} 
processes, which are not included in the Geant4.9.2 package.

Since a good agreement between the data and two models was achieved in the 
shape analysis, we extracted an average visible nuclear recoil energy, 
E$_{vr}$, from each bin using the difference between the measured visible 
energy and the $^{72}$Ge$^{*}$(0$^{+}$) to $^{72}$Ge(0$^{+}$) transition energy, 
691.6 keV. The corresponding nuclear recoil energy, E$_{r}$, was determined 
utilizing the Monte Carlo simulation. Table~\ref{tab:table2} displays the 
extracted results.
\begin{table}[tb!!]
\caption{The extracted average visible nuclear recoil energy from the data and the corresponding
nuclear recoil energy from the Monte Carlo simulation. E$_{vr}$ contains a statistical error of 14\% per energy bin and a systematic 
error of 4.6 keV added in quadrature. There are no errors assigned to the nuclear recoil energy obtained from the Monte Carlo simulation. }
\label{tab:table2}
\begin{tabular}{|c|c|c|c|c|c|c|c|}
\hline \hline
 E$_{vr}$ (keV) & 0.5$\pm$ 0.5& 1.4$\pm$1.4 & 4.4$\pm$4.4 & 10.4$\pm$4.8 & 16.4$\pm$5.1 & 22.4$\pm$5.6 & 28.4$\pm$6.1\\
\hline 
 E$_{r}$ (keV)  & 2.7 &  6.7  & 17.7 & 37.3  & 56.6 & 76.3 & 97.3\\
\hline \hline
\end{tabular}
\end{table}

\begin{figure}[tb!!]
\includegraphics[angle=0,width=12.cm] {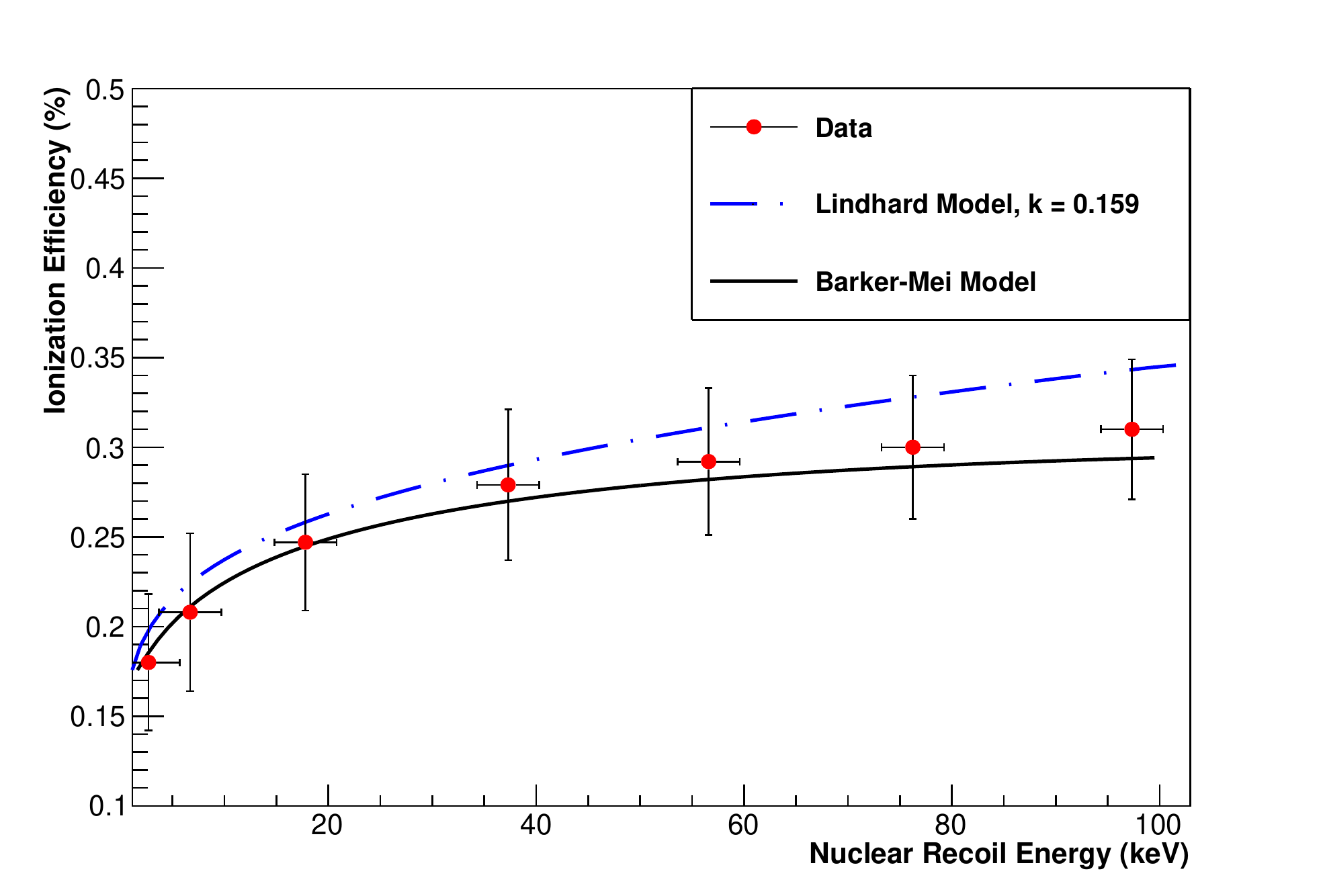}
\caption{\small{Extracted ionization efficiency from the shape analysis. The error bars account for a 
statistic error of 14\% per bin and the variation due to the systematic error of 4.6 keV added in quadrature.}}
\label{fig:efficiency}
\end{figure}

Using $\frac{E_{vr}}{E_{r}}$, Fig.~\ref{fig:efficiency} shows the extracted 
ionization efficiency from the data and Monte Carlo simulation. Note that this 
is not a direct measurement of ionization efficiency, but an extraction using
the shape analysis.

\section{Conclusion}
We took measurements using an AmBe neutron source, incident on a germanium 
detector, for a total of 35 days. The characteristic quasi-triangular shape 
located at 691.6 keV represents the E$_0$ transition of germanium-72.  All 
peaks in the region of interest have been identified, thereby confirming source 
related and environmental backgrounds. The unique quasi-triangular shape 
induced by neutrons can be described using the Moyal distribution with the 
fitted parameters $R$ = 0.06 $\pm$ 0.01 (stat) and $E_{mpv}$ = 696 $\pm$ 4.6 
(sys) $\pm$ 1.4 (stat) keV, after background subtraction. Utilizing these 
parameters, we derived the most probable value for nuclear recoils as 4.4 $\pm$ 
0.007 (sys) $\pm$ 0.002 (stat) keV electronic equivalent energy, which 
corresponds to a nuclear recoil energy of 17.5 $\pm$ 0.12 (sys) $\pm$ 0.035 
(stat) keV. The average path length in the germanium detector for 17.5 keV 
nuclear recoils was approximately 0.014 $\mu$m.

A Monte Carlo simulation employing a corrected Geant4.9.2 package was modified 
to duplicate the same experimental setup and AmBe neutron source. The Lindhard 
($k$ = 0.159) and Barker-Mei models were used to apply ionization efficiency to 
the energy spectrum and were compared to the experimental measurements. A 
bin-to-bin shape analysis was performed, and the difference between the 
measurements and two models were calculated. We obtained a percent difference 
that was less than 4\% for the collected data and two Monte Carlo simulations 
(see Table 1). Using the shape analysis, we calculated 
the most probable values for visible nuclear recoils (as shown in Table 1), 
extracted the average visible nuclear recoil energy (see Table 2), and obtained the 
corresponding nuclear recoil energy from the Monte Carlo simulation (see Table 2). 
The extracted average ionization efficiency is shown in 
Fig.~\ref{fig:efficiency}. These values are in agreement with the Lindhard and Barker-Mei 
models in the energy range of 1 to 100 keV. Therefore, the Lindhard model 
(with $k$ = 0.159) and Barker-Mei model can be used to determine the 
ionization efficiency in germanium detectors for 1 to 100 keV nuclear recoil 
energy.

\section*{Acknowledgments}
The authors wish to thank Iseley Marshall and Angela A. Chiller for their
 careful reading of this manuscript. Additionally,
the authors 
would like to thank Rupak Mahapatra for his comments and suggestions.  
This work was supported in part by NSF PHY-0758120, DOE grant DE-FG02-10ER46709, 
the Office of Research at the University 
of South Dakota and a 2010 research center support by the State of South Dakota. 

%
%

\end{document}